\documentstyle[aps,prd]{revtex}
\begin{document}
\title{Scale anomalies imply violation of the averaged null energy condition}
\author{Matt Visser\cite{e-mail}}
\address{Physics Department, Washington University, St. Louis,
         Missouri 63130-4899}
\date{25 August 1994}
\maketitle
\begin{abstract}
Considerable interest has recently been expressed regarding the
issue of whether or not quantum field theory on a fixed but curved
background spacetime satisfies the averaged null energy condition
(ANEC). A comment by Wald and Yurtsever [Phys. Rev. {\bf D43}, 403
(1991)] indicates that in general the answer is no.  In this note
I explore this issue in more detail, and succeed in characterizing
a broad class of spacetimes in which the ANEC is guaranteed to be
violated.  Finally, I add some comments regarding ANEC violation
in Schwarzschild spacetime.
\end{abstract}

\section{INTRODUCTION}
\newtheorem{theorem}{Theorem}
\newtheorem{definition}{Definition}

Considerable interest has recently been expressed regarding the
issue of whether or not quantum field theory on a fixed but curved
background spacetime satisfies the averaged null energy condition
(ANEC).  The ANEC is one of the weakest energy conditions in common
use --- it is used for instance as a hypothesis in proving sundry
focussing theorems for null geodesics~\cite{Borde87}, in proving
certain modifications of the original singularity theorems~\cite{Roman88},
in proving the topological censorship
theorem~\cite{Topological-Censorship}, and in proving a recent
version of the positive mass theorem~\cite{Penrose-Sorkin-Woolgar}.
It is thus a topic of considerable interest to know if the ANEC can
itself be proved from more some basic postulates of quantum field
theory. Promising progress toward such a proof was made by
Klinkhammer~\cite{Klinkhammer91}, by Yurtsever~\cite{Yurtsever90a},
and by Wald and Yurtsever~\cite{Wald-Yurtsever}. Unfortunately, it
is now realised that in general the ANEC  must fail. See the ``note
added in proof'' on page 415 of~\cite{Wald-Yurtsever}. The Wald-Yurtsever
``note added in proof'' applies only to generic perturbations around
a flat Minkowski spacetime.  In this letter I wish to explore this
issue in a little more depth and detail.

I show that in a general spacetime the relevant issue is whether
or not the renormalized expectation value of the quantum stress--energy
tensor possesses a scale anomaly. A scale transformation is defined
as a position independent conformal rescaling of the metric.  Scale
anomalies are related to conformal anomalies but are considerably
easier to deal with. I then analyze conditions under which the
scale anomaly vanishes: The scale anomaly is zero for Einstein
spacetimes (in particular for Schwarzschild spacetime) and is zero
for conformally flat spacetimes, but is non--zero generically.

Some more detailed statements about the status of the ANEC in
Schwarzschild spacetime can be made by appealing to Howard's
numerical calculations~\cite{Howard,Howard-Candelas} of the
renormalized vacuum expectation value of the stress--energy tensor.

\section{Averaged null energy condition}

\begin{definition}
The averaged null energy condition is said
to hold on a null curve $\gamma$ if
\begin{equation}
\int_\gamma T_{\mu\nu} \;  k^\mu k^\nu \; d\lambda \geq 0.
\end{equation}
Here $\lambda$ is a generalized affine parameter for the null curve,
the tangent vector being denoted by $k^\mu$.
\end{definition}
Technical points: (1) If $\gamma$ is a null geodesic then the
generalized affine parameter specializes to the ordinary affine
parameter. See for example,~\cite{Hawking-Ellis}, page 259.  (2)
Permitting arbitrary parameterizations would not be useful. If
arbitrary parameterizations were to be allowed, the ANEC would be
equivalent to the ordinary null energy condition.

To get a physical feel for what is going on, suppose, for the sake
of discussion, that the stress--energy tensor is type I.  In a
suitable orthonormal frame, the energy density and three principal
pressures are given by
\begin{equation}
T^{\hat\mu\hat\nu} =
\left[ \matrix{\rho&0&0&0\cr
               0&p_1&0&0\cr
               0&0&p_2&0\cr
	       0&0&0&p_3\cr } \right].
\end{equation}
(See, for instance~\cite{Hawking-Ellis}, page 89.) In this orthonormal
frame one can define a function $\xi$, and direction cosines
$\cos\psi_j$, by
\begin{equation}
k^{\hat \mu} = \xi \; (1;\cos\psi_j).
\end{equation}
Then
\begin{equation}
ANEC \iff
\int_\gamma \big[ \rho+ \sum_j (\cos^2\psi_j) \; p_j
            \big] \;
\xi^2 \; d\lambda \geq 0.
\end{equation}
In applications one typically requires the ANEC to hold on some
suitable class $\Gamma=\{\gamma\}$ of inextendible null geodesics.

\section{ANEC violations}

The central result of this letter is
\begin{theorem}
In any $(3+1)$--dimensional spacetime, for any conformally coupled
quantum field, in any conformal quantum state:  If the scale anomaly
is non--zero, then the renormalization scale $\mu$ can be chosen
in such a way that the ANEC is violated.
\end{theorem}
As background, recall the tremendous amount of technical machinery
available for analyzing the behaviour of the stress--energy tensor
under conformal deformations~\cite{Birrell-Davies,Fulling,Page82}.

A {\em scale transformation} is just the special case of a conformal
transformation when the conformal rescaling factor is {\em position
independent}
\begin{equation}
g(x) \to \bar g(x) = \Omega^2 g(x).
\end{equation}
For a conformally coupled quantum field, consider the renormalized
vacuum expectation value of the quantum stress--energy tensor
evaluated on a conformal quantum state. In $(3+1)$--dimensions this
object undergoes an anomalous scaling transformation
\begin{equation}
T^\mu{}_\nu(\bar g) =
\Omega^{-4} \left(
T^\mu{}_\nu( g) - 8 a \ln\Omega
\left[\nabla_\alpha \nabla^\beta + {1\over2} R_\alpha{}^\beta \right]
C^{\alpha\mu}{}_{\beta\nu} \right).
\end{equation}
This is a standard result. For example, this is a special case of
equation (66) of Page~\cite{Page82}. For future convenience it is
useful to define
\begin{equation}
Z^\mu{}_\nu \equiv
\left[\nabla_\alpha \nabla^\beta + {1\over2} R_\alpha{}^\beta \right]
C^{\alpha\mu}{}_{\beta\nu} .
\end{equation}

The anomalous scaling of the stress--energy under a rescaling of
the metric is a consequence of the fact that in regularizing the
conformal quantum field one has had to introduce a cutoff. This
cutoff breaks the conformal invariance and, after proper renormalization
to remove explicit cutoff dependence, results in a dimensional
transmutation effect whereby the expectation value depends on a
so--called renormalization scale $\mu$. (See, for
example,~\cite{Page82,Horowitz80}).  One may make this renormalization
scale dependence explicit by writing
\begin{equation}
T^\mu{}_\nu(\Omega g; \mu ) = T^\mu{}_\nu(g; \mu/\Omega ).
\end{equation}
Thus, the scale anomaly may be recast as
\begin{equation}
T^\mu{}_\nu(g; \mu/\Omega) =
\Omega^{-4}
\left[ T^\mu{}_\nu(g; \mu) - 8 a \ln\Omega \;  Z^\mu{}_\nu  \right].
\end{equation}
The coefficient $a$ is exactly the same as that arising in
the perhaps more familiar conformal trace anomaly
\begin{eqnarray}
T^\mu{}_\mu &=&
a \; (C_{\mu\nu\lambda\rho} C^{\mu\nu\lambda\rho})
+ b \; (*R_{\mu\nu\lambda\rho} *R^{\mu\nu\lambda\rho})
\nonumber\\
&&+ c \; \Box R
+ d R^2.
\end{eqnarray}
Now consider the effect of the scale anomaly on the ANEC integral.
Let $\gamma$ be any null curve of the metric $g$ parameterized
by a generalized affine parameter $\lambda$. Then
\begin{eqnarray}
I_\gamma(\mu/\Omega) &\equiv&
\int_\gamma T^\mu{}_\nu(g;\mu/\Omega) \; k_\mu k^\nu \; d\lambda,\\
&=&
\int_\gamma \Omega^{-4}
\left[
T^\mu{}_\nu(g; \mu) - 8 a \ln\Omega Z^\mu{}_\nu
\right] k_\mu k^\nu d\lambda,
\nonumber\\
&=&
\Omega^{-4} I_\gamma(\mu)
- 8 a \Omega^{-4} \ln\Omega \; \int_\gamma
Z^\mu{}_\nu \; k_\mu k^\nu \; d\lambda.
\end{eqnarray}
For convenience define
\begin{equation}
J_\gamma =
\int_\gamma Z^\mu{}_\nu \; k_\mu k^\nu \; d\lambda.
\end{equation}
Now if $J_\gamma \neq 0$ it is always possible to choose $\ln\Omega$
sufficiently large (either positive or negative) to force $I_\gamma
<0$. This means that ANEC is guaranteed to be violated for the null
curve $\gamma$ for a suitable choice of the renormalization scale
$\mu$. [Technical points: (1) For simplicity I have assumed that
all the integrals converge.  The analysis of Wald and
Yurtsever~\cite{Wald-Yurtsever}, pages 404--405, can be adapted to
deal with the case where the ANEC integral does not converge. (2)
If $Z^\mu{}_\nu \neq 0$ then a generic null curve which intersects
the support of $Z^\mu{}_\nu$ will have $J_\gamma \neq 0$.] This
completes the proof.

An alternative formulation is
\begin{theorem}
In $(3+1)$--dimensional spacetime the ANEC is guaranteed to be
violated whenever
\begin{equation}
Z^\mu{}_\nu \equiv
\left[\nabla_\alpha \nabla^\beta + {1\over2} R^\alpha{}_\beta \right]
C^{\alpha\mu}{}_{\beta\nu}  \neq 0.
\end{equation}
\end{theorem}
Comments:
\begin{itemize}
\item
If $Z^\mu{}_\nu = 0$ we cannot deduce that the ANEC {\em must} be
satisfied.  We can only deduce that it is {\em possible} for the
ANEC to be satisfied.  Cases are known where $Z^\mu{}_\nu=0$, with
the ANEC being satisfied on null geodesics, but with the ANEC
violated along certain classes of non--geodesic null
curves~\cite{Klinkhammer91}. For another example, see the discussion
of the Schwarzschild spacetime given below.
\item
In Minkowski spacetime $Z^\mu{}_\nu = 0$. Thus Minkowski spacetime
evades the no--go theorem. This result is compatible with the
theorems proved by Klinkhammer~\cite{Klinkhammer91}.
\item
In conformally flat spacetimes $Z^\mu{}_\nu = 0$. All conformally
flat spacetimes evade the no--go theorem.
\item
In $(1+1)$--dimensions the scale anomaly vanishes. [For example,
take equation (6.134) of Birrell and Davies~\cite{Birrell-Davies}
and set $\nabla\Omega=0$.] This happy accident is due to the fact
that all two--dimensional manifolds are conformally flat. Thus all
$(1+1)$--dimensional spacetimes evade the no--go theorem.  This
result is compatible with the theorems proved by Yurtsever
\cite{Yurtsever90a} and Wald and Yurtsever~\cite{Wald-Yurtsever}.
\item
In any Einstein spacetime $Z^\mu{}_\nu = 0$.  To see this, recall that
an Einstein spacetime is defined by
\begin{equation}
R_{\mu\nu} = \Lambda g_{\mu\nu}.
\end{equation}
Then note
\begin{equation}
R_\alpha{}^\beta  C^{\alpha\mu}{}_{\beta\nu} =
\Lambda g_\alpha{}^\beta  C^{\alpha\mu}{}_{\beta\nu} = 0.
\end{equation}
Furthermore, starting from the Bianchi identities, a single
contraction yields (See, for example,~\cite{Hawking-Ellis}, equation
(2.32) page 43.)
\begin{equation}
\nabla_\mu C^\mu{}_{\nu\sigma\rho} =
-R_{\mu[\sigma;\rho]} + {1\over6} g_{\mu[\sigma} R_{;\rho]}
\end{equation}
In view of the definition of an Einstein spacetime this implies
\begin{equation}
\nabla_\mu C^\mu{}_{\nu\sigma\rho} = 0.
\end{equation}
Therefore $Z^\mu{}_\nu = 0$; all Einstein spacetimes have zero
scale anomaly and evade the no--go theorem. In particular,
Schwarzschild spacetime has $Z^\mu{}_\nu = 0$ and so evades the
no-go theorem.
\item
On the other hand, generic perturbations of any spacetime with zero
scale anomaly will lead to a non--zero scale anomaly.
So generic perturbations of manifolds satisfying the ANEC lead to
manifolds where the ANEC is violated by a suitable choice of
renormalization scale. In this sense violations of the ANEC are
generic. The observation generalizes the case of linearized
perturbations around flat Minkowski space which was addressed
in~\cite{Wald-Yurtsever}.
\item
If one chooses to work with massless quantum fields that are not
conformally coupled the analysis is more complicated --- this is
why the Wald--Yurtsever involves {\em two} independent fourth order
curvature tensors~\cite{Wald-Yurtsever}. (Conformal coupling implies
that their coefficient $b$ is zero.)
\end{itemize}

\section{Schwarzschild spacetime}

In the particular case of Schwarzschild spacetime the explicit
numerical calculations of Howard~\cite{Howard}, and Howard
and Candelas~\cite{Howard-Candelas} permit one to make some more
detailed statements about the ANEC.
By spherical symmetry one knows that
\begin{equation}
\langle 0_H | T^{\hat\mu\hat\nu} | 0_H \rangle \equiv
\left[ \matrix{\rho&0&0&0\cr
               0&-\tau&0&0\cr
               0&0&p&0\cr
	       0&0&0&p\cr } \right].
\end{equation}
For example, at the horizon, a conformally coupled scalar field in
the Hartle--Hawking vacuum state has stress--energy\begin{equation}
\langle 0_H | T^{\hat\mu\hat\nu} | 0_H \rangle \approx
{\pi^2\over 90} \left({1\over 8\pi m}\right)^4
\left[ \matrix{-37.728&0&0&0\cr
               0&+37.728&0&0\cr
               0&0&+10.29&0\cr
	       0&0&0&+10.29\cr } \right].
\end{equation}
This immediately implies, from the negativity of $\rho$, violation
of the weak and dominant energy conditions.  Checking that $\rho
+ p <0$, one also sees that the null and strong energy conditions
are violated. (For definitions, see for instance~\cite{Hawking-Ellis}).
Warning:  The paper by Fawcett~\cite{Fawcett} unfortunately contains
an error. See~\cite{Howard,Howard-Candelas}.

The status of the ANEC is considerably more complicated: rewrite
the ANEC integral  as
\begin{eqnarray}
I_\gamma
&=&
\int_\Gamma (\rho -\tau \cos^2\psi + p \sin^2\psi) \; \xi^2 \; d\lambda,
\\
&=&
\int_\Gamma ([\rho -\tau]  + [\tau +p ]\sin^2\psi) \; \xi^2 \; d\lambda.
\end{eqnarray}
Comments:
\begin{itemize}
\item
The numerical data of Howard~\cite{Howard} indicate that outside
the horizon $\rho -\tau >0$. ($\rho = \tau$ at the horizon.) The
numerical data covers the range from $r=2\, m$ to $r=6.8\, m$. For
larger radial distances one resorts to Page's analytic approximation,
which becomes increasingly good as one moves further away from the
horizon~\cite{Howard,Page82}. This implies that the ANEC is satisfied
for radial null geodesics. Caveat: The ANEC integral has to be
modified to a one--sided integral by cutting it off at the event
horizon.
\item
Null geodesics that come in from infinity and return to infinity
never get closer to the origin than $r=3m$. See, for instance,~\cite{MTW},
pages 672--678. Inspection of Howard's numerical data indicates
that the integrand of the ANEC integral is strictly positive for
$r \geq 3\,m$. (Inspection shows that $\rho$, $-\tau$, and $p$ are
all positive for $r \geq 3\,m$.) Thus the ANEC is satisfied along
all null geodesics that come from, and return to, infinity.
\item
For the unstable circular photon orbit at $r=3m$ the ANEC integral
is proportional to $\rho+p$. Inspection of Howard's numerical data
indicates that $\rho+p > 0$ at $r=3m$, so the ANEC is satisfied
along this particular null geodesic.
\item
For incoming null geodesics with smaller than critical impact
parameter, the null geodesic may circle the black hole a large
number of times, but is guaranteed to ultimately plunge into the
event horizon~\cite[pages 672--678]{MTW}. This makes the analysis
a little more subtle. Inspect the ANEC integrand. Inspecting the
data shows that $\rho-\tau$ is always positive, while $\tau + p$
is always negative.  (Where the numerical data is available, check
this using the numerical data. At large radius, check this using
Page's approximation.) Now use the fact that for an infalling null
geodesic
\begin{equation}
\sin^2 \psi < 27 {m^2\over r^2} (1- 2m/r).
\end{equation}
Since $\tau + p$ is negative, this implies a lower bound on the
ANEC integral
\begin{eqnarray}
I_\gamma \geq
\int_\Gamma \big\{[\rho -\tau] +
           27 [\tau +p ] {m^2 \over r^2} (1- 2m/r)\big\}
	   \; \xi^2 \; d\lambda.
\end{eqnarray}
Inspection, either of the numerical data, or of Page's approximation,
indicates that the integrand of this lower bound is strictly positive
(zero at the horizon). Thus the ANEC holds on all infalling null
geodesics.
\item
By time reversal, the ANEC holds on all outgoing null geodesics
that reach infinity.
\item
There remains the issue of those null geodesics that emerge from
the horizon, circle the black hole, and plunge back in. For such
trapped null geodesics one has at all times
\begin{equation}
\sin^2 \psi > 27 {m^2\over r^2} (1- 2m/r).
\end{equation}
Unfortunately, this observation does not permit one to derive a
useful bound on the ANEC integral. (By working in terms of the
``impact parameter'' it is relatively easy to convince oneself that
the ANEC integrand is permitted to be negative for regions sufficiently
close to the event horizon.)
\item
For any circular null curve at fixed $r$ ({\em not} a geodesic
except in the case of $r=3m$)  the ANEC integral is still proportional
to $\rho+p$. Inspection of Howard's numerical data indicates that
$\rho+p < 0$ for  $r \lesssim 2.25\, m$. (Page's analytic approximation
gives a slightly different result, $\rho+p < 0$ for  $r < 2.18994\,
m$.) Thus the ANEC is {\em not} satisfied for this particular class
of non--geodesic null curves.
\end{itemize}
Collecting these comments, I have:
\begin{theorem}
Consider the renormalized vacuum expectation value of the stress--energy
tensor for a conformally coupled scalar field in Schwarzschild
spacetime. Then (1) the ANEC is satisfied for all null geodesics
that reach spatial infinity. (2) there are non--geodesic null curves
along which the ANEC is violated.
\end{theorem}

It should be possible to generalize these observations. For instance:
(1) The behaviour of the ANEC on trapped null geodesics is still
somewhat obscure.  Presumably the ANEC depends on the ``impact
parameter'' in an interesting way.  (2) By using Page's analytic
approximation one could write down analytic estimates for the ANEC
integral. I have not attempted to do so in this letter because I
felt it to be instructive to first develop simple physical bounds
on the ANEC integral. (3) It would be nice to go beyond the numerics;
to develop some exact analytic arguments that go beyond the Page
approximation.

\section{Discussion}

Investigation of the properties of the averaged null energy condition
is of considerable interest to diverse applications in both classical
and semiclassical quantum gravity. It is now clear that, in general,
semiclassical quantum fields do {\em not} satisfy the ANEC. In
this letter, I have related ANEC violations to the existence of
the scale anomaly --- If the scale anomaly is non--zero then the
ANEC is guaranteed to be violated. Even if the scale anomaly
vanishes, this does not necessarily imply that the ANEC is satisfied:
one has to do a case by case analysis. As an example, the situation
in Schwarzschild spacetime was investigated.

\acknowledgements

This research was supported by the U.S. Department of Energy.


\end{document}